\documentclass{article}

\usepackage{PRIMEarxiv}

\usepackage[utf8]{inputenc} % allow utf-8 input
\usepackage[T1]{fontenc}    % use 8-bit T1 fonts
\usepackage{hyperref}       % hyperlinks
\usepackage{url}            % simple URL typesetting
\usepackage{booktabs}       % professional-quality tables
\usepackage{amsfonts}       % blackboard math symbols
\usepackage{nicefrac}       % compact symbols for 1/2, etc.
\usepackage{microtype}      % microtypography
\usepackage{lipsum}
\usepackage{fancyhdr}       % header
\usepackage{graphicx}       % graphics
\graphicspath{{media/}}     % organize your images and other figures under media/ folder
\usepackage{amsmath,amssymb,amsfonts}

\title{A template for Arxiv Style
\thanks{\textit{\underline{Citation}}: 
\textbf{Authors. Title. Pages.... DOI:000000/11111.}} 
}

\author{
  Vaibhav Gupta \\
 Data Engineering\\
  Helmut-Schmidt-University \\
  Hamburg\\
  \texttt{guptav@hsu-hh.de} \\
  \And
  Florian Grensing \\
  Data Engineering\\
  Helmut-Schmidt-University \\
  Hamburg\\
   \And
  Beyza Cinar \\
  Data Engineering\\
  Helmut-Schmidt-University \\
  Hamburg\\
  \And
   Louisa van den Boom \\
  Clinic for Pediatric and Adolescent Medicine\\
  Helios Klinikum Gifhorn GmbH\\ 
  Gifhorn, Germany \\
  \And
   Maria Maleshkova \\
  Data Engineering\\
  Helmut-Schmidt-University \\
  Hamburg\\ 
}

\title{\LARGE \bf
Towards Improved Short-term Hypoglycemia Prediction and  Diabetes Management based on Refined Heart Rate Data 
}
 
\begin{document}

\maketitle
\thispagestyle{empty}
\pagestyle{empty}

%%%%%%%%%%%%%%%%%%%%%%%%%%%%%%%%%%%%%%%%%%%%%%%%%%%%%%%%%%%%%%%%%%%%
\begin{abstract}
Hypoglycemia is a severe condition of decreased blood glucose, specifically below 70 mg/dL (3.9 mmol/L). This condition can often be asymptomatic and challenging to predict in individuals with type 1 diabetes (T1D). Research on hypoglycemic prediction typically uses a combination of blood glucose readings and heart rate data to predict hypoglycemic events. Given that these features are collected through wearable sensors, they can sometimes have missing values, necessitating efficient imputation methods. This work makes significant contributions to the current state of the art by introducing two novel imputation techniques for imputing heart rate values over short-term horizons: Controlled Weighted Rational Bézier Curves (CRBC) and Controlled Piecewise Cubic Hermite Interpolating Polynomial with mapped peaks and valleys of Control Points (CMPV). In addition to these imputation methods, we employ two metrics to capture data patterns, alongside a combined metric that integrates the strengths of both individual metrics with RMSE scores for a comprehensive evaluation of the imputation techniques. According to our combined metric assessment, CMPV outperforms the alternatives with an average score of 0.33 across all time gaps, while CRBC follows with a score of 0.48. These findings clearly demonstrate the effectiveness of the proposed imputation methods in accurately filling in missing heart rate values. Moreover, this study facilitates the detection of abnormal physiological signals, enabling the implementation of early preventive measures for more accurate diagnosis.
\end{abstract}

\section{Introduction}

Type 1 diabetes (T1D) is a chronic metabolic disorder characterised by the autoimmune distruction of pancreatic beta-cells, leading to insufficient insulin production. To mitigate elevated glucose levels, patients require insulin therapy, which poses a risk of hypoglycemia when blood glucose levels fall below 70 mg/dL. Hypoglycemia can be life-threatening and is often asymptomatic, therefore, early detection and intervention are necessary~\cite{ADA2023}. FDA-approved wearable devices, integrated with machine learning (ML) algorithms, can prevent severe events such as hypoglycemia through real-time monitoring and alerts~\cite{Nomura2021,Cinar2025DiaData, Grensing2025EarlyWarning}. Incorporating multivariate data inputs enhances personalisation and improves predictive model accuracy~\cite{Site2023}. Specifically, electrical sensors such as photoplethysmography (PPG) and electrocardiography (ECG) contribute to the robustness of data-driven models. In particular, ECG data can be used to predict hypoglycemia independently of the continuous glucose monitoring (CGM) inputs, underscoring its predictive capacity and its relationship to glycemic variability~\cite{Dave_Noninvasive_2022}. Heart rate (HR) is frequently utilized as input for ML models, demonstrating a significant correlation with glucose variability and hypoglycemic episodes~\cite{Site2023,Weiler2017, Yu2006}. 
 
Wearable devices enable continuous, non-invasive data collection but face challenges such as sensor limitations, environmental factors, and user-related issues. These can lead to unreliable data due to artefacts, noise, outliers, and data gaps, requiring preprocessing techniques, including imputation, to enhance quality and model efficiency~\cite{diadata_kaggle}. Various methods can be employed to address missing data. Analysis of multiple studies utilizing imputation techniques~\cite{Benchekroun2023, Mochurad2023ParallelAF,vahedi} reveals statistical models are typically used for short-time horizons, while linear interpolation is effective for gaps of up to two hours~\cite{SuwenLin, Bertachi2020}. For longer gaps of more than two hours, regression models/deep learning techniques~\cite{SuwenLin} or aggregation techniques are usually applied~\cite{Bertachi2020}. For instance, Leutheuser et al. impute by using the mean values (aggregation) of the same day~\cite{Leutheuser2024}. However, statistical methods usually cannot capture the natural pattern of the HR or ECG signals, limiting their ability to effectively replicate natural clinically relevant signals. In particular, for the imputation of HR signals, linear imputation and Piecewise Cubic Hermite Interpolating Polynomial (PCHIP) are popular for short-term gaps~\cite{Benchekroun2023,Leutheuser2024}. Regression models are primarily used for longer gaps because they train on available datasets to predict missing values, but their computational expense can make them less suitable for shorter gaps~\cite{computers13120312}. Furthermore, imputation techniques, including statistical and regression models, can also be seen as a way to simulate physiological signals by predicting trends from real data. This makes them strong candidates for developing realistic simulators for physiological signals.
 
Here we address the limitations of statistical techniques in effectively imputing short-term gaps, thus the detection of extreme physiological conditions. The primary focus of our work is \textbf{twofold}: 1) First, it aims to enhance the quality of HR data based on imputing missing values for time intervals up-to 30 mins, ensuring its reliability for clinical applications based on  data analysis. 2) Second, due to natural heart rate fluctuations, it addresses the need to introduce a multidimensional evaluation framework to assess imputed data fidelity relative to original values, ensuring the comprehensive assessment of data quality.

To achieve this, we make the following \textbf{key contributions:} 1) We introduce \textbf{two novel imputation methods} that capture the natural dynamics of physiological signals, significantly enhancing the accuracy and reliability of data imputation. 2) Additionally, we propose  \textbf{two novel metrics} for capturing and evaluating the fluctuating patterns of imputed HR values, along with a combined evaluation metric. 3) Finally, we present a preliminary framework for developing a \textbf{heart rate simulator}, laying the foundation for future advancements in HR data synthesis. 
 
This study analyses a critical yet underexplored challenge in handling missing data within physiological signals, with a particular focus on best practices for imputing HR data over short-time horizons. 
By establishing a foundational baseline for developing solutions capable of imputing HR values, this study paves the way for potential applications in predicting various cardiovascular and metabolic conditions associated with heart rate fluctuations. The remainder of this paper is structured as follows. Section \ref{sec:LiteratureReview} reports on the state of the art of imputation methods on HR signals. Section \ref{sec:Methods} introduces the used dataset, the proposed and applied imputation methods and metrics. The obtained results are reported in section \ref{sec:Results} and discussed in section \ref{sec:Discussion}. Finally, key findings are summarized in section \ref{sec:Conclusion}.

\section{Literature Review}
\label{sec:LiteratureReview}

The application of imputation techniques for handling missing values in vital parameters has increased considerably in recent years. The studies reviewed are categorized based on two criteria: 1) studies utilising imputation techniques for missing ECG and HR features for hypoglycemia prediction and 2) studies, primarily focusing on imputing missing values of ECG and HR features.
Studies belonging to the first criterion are limited. Mantena \cite{Mantena_2022} utilises KNN-based imputation methods, Vahedi et al. \cite{vahedi} and Leutheuser et al. \cite{Leutheuser2024} substitute with mean values for HR feature imputation, and Bertachi et al. \cite{Bertachi2020} used linear interpolation for a maximum of 2 h and removed remaining gaps. Dave et al. \cite{Dave_Noninvasive_2022} impute any missing data points in the ECG segment of one minute by averaging the feature in the segment. Hypoglycemic prediction studies focus on the prediction outcomes but rarely emphasise the effectiveness of the imputation techniques employed. Multiple research studies highlight various imputation procedures for mitigating data gaps to improve prediction outcomes.

Studies that meet the second criterion include Lin et al. \cite{SuwenLin} and Mochurad et al. \cite{Mochurad2023ParallelAF}, who develop certain methods for computing and imputing gaps. Lin et al. \cite{SuwenLin} generated gap sizes of 2 h, 4 h, 6 h, and 8 h within a 24-h timeframe, whereas Mochurad et al. \cite{Mochurad2023ParallelAF} generated random gaps constituting 20 per cent of the entire dataset. Lin et al. \cite{SuwenLin} uses traditional imputation methods (linear interpolation, exponentially weighted moving average, K-nearest neighbors, Kalman smoothing, and Last Observation Carried Forward) alongside deep learning techniques (Denoising Autoencoder, bi-directional RNN, Context Encoders, Spatial-Temporal Completion Network for Video Inpainting and HeartImp). HeartImp is reported to be the best imputation method for all gap sizes achieving RMSE, MAPE and MAE of 14.37, 10.58, 0.13 for Garmin Data (2 h for June) and 15.56, 10.18 and 0.12 for Fitbit data (2 h for July), respectively. Mochurad et al. \cite{Mochurad2023ParallelAF} employ Bézier and B-spline interpolation for gap imputation, utilising MAPE and $R^{2}$ as assessment metrics. Bézier yields MAE and $R^{2}$ scores of 0.09 and 0.93, respectively, whereas B-Spline interpolation produces scores of 0.19 and 0.87.

Despite advancements in imputing features for ECG and HR values in wearable sensor datasets for hypoglycemia prediction, there is still a lack of effective imputation methods specifically for short interval gaps (under 30 minutes). To the best of our knowledge, no study has focused on imputation techniques tailored to different short-interval gap sizes in hypoglycemia prediction datasets. Therefore, this study proposes two new imputation methods, specifically tailored to HR values for short term horizons. Furthermore, current research typically evaluates imputation techniques using metrics such as RMSE, MAPE, and MAE, which measure the numerical deviation between imputed and original values. However, these metrics fail to adequately assess the accuracy of imputation methods, particularly regarding the preservation of the feature shape of the original data \cite{Boursalie2022}. To address this gap, we have introduced two shape-preserving metrics that measure and assess the density and peak distribution of imputed HR values while maintaining correspondence with the original HR data.

\section{Methods and Research Design}
\label{sec:Methods}

This study aims to efficiently impute missing HR values to improve the performance of hypoglycemia prediction methods for short-term horizons (under 30 minutes). The methodology of the imputation takes the architecture from Impute Paradigm \cite{GUPTA2025}, in which different gaps are imputed with different imputation techniques. Therefore, this study focuses on imputing short-term gaps, while the application of similar techniques to long-term gaps remains unexplored. To address key research gaps in the state-of-the-art, this study introduces two imputation techniques for handling missing HR values in the dataset to improve hypoglycemia prediction. These techniques include Controlled Weighted Rational Bézier Curves (CRBC) and Controlled PCHIP, which leverage Mapped Peaks and Valleys of Control Points (CMPV) to preserve the shape and structure of the data. To assess the effectiveness of the imputation techniques, we intentionally create gaps in the data, allowing for a comparison between actual and predicted sequences. The effectiveness of the proposed methods are compared with state of art imputation techniques. To evaluate the performance of the methods, one state of the art evaluation metric and two proposed evaluation metrics which efficiently captures the pattern of the data are proposed. Finally, the results are evaluated by the combined aggregated metric. Graphs illustrating the imputation methods are presented to substantiate the metric results. In the following subsection, we will examine the structure of the design in detail.

\subsection{ \textbf{Dataset Description}}
This study evaluates the best imputation technique for the HR feature of the D1NMAO dataset~\cite{DUBOSSON201892} over short-term horizons. This dataset meets the criteria of our twofold objective, which we want to accomplish through our study. Data of 9 patients with T1D has been utilised for our experiments. The dataset comprises of continuous glucose data recorded every 5 minutes using the iPro2 Professional CGM sensor. ECG, breathing and accelerometer outputs measurements are collected every second using the Zephyr Bioharness 3 sensor. These raw signals can be used directly and the device also computes additional aggregated metrics such as HR, Breathing Rate (BR), activity level, posture, etc. Further insights of the datasets are presented in table \ref{table1}. 

\begin{table}
\centering
\caption{Overview of D1NAMO dataset }
\label{table1}
\renewcommand{\arraystretch}{1.5} 
\setlength{\tabcolsep}{0.5pt}
\begin{tabular}{|p{105pt}|p{125pt}|}
\hline 
\textbf{ Dataset} & \textbf{ D1NAMO}  \\ 
\hline
\textbf{ Subjects} &  20 healthy, 9 T1D patients \\
\hline
\textbf{ CGM Device} &  iPro2 Professional CGM sensor \\
\hline
\textbf{ Physiological Sensor}  & Zephyr Bioharness 3 sensor\\
\hline 
\textbf{ Signals recording}  &  450h\\
 \hline
\textbf{ Glucose measurements} & 8414\\
 \hline
\textbf{ Food pictures}	& 106 \\
 \hline
\end{tabular}
\end{table}

\subsection {\textbf{Preparing the Experimental Data}} 

To enable an effective evaluation, gaps were deliberately created and 
the removed sequence was used as ground truth. For this, we combined all HR files for each individual patient within the dataset. Following this, we identified all areas which are sufficiently long without gaps. Once we found these areas, we created one random gap for every four times the gap size. These are chosen at random, with the condition that there are no missing values for at least half the size of the gap before and after the gap, as these are required for the imputation. We also ensure that there is no overlap in the gaps created. Fewer gaps could be created for larger gap sizes, as these require more data.

\subsection{\textbf{Controlled Weighted Rational Bézier Curves (CRBC)}}

Chutchavong et al. \cite{Chutchavong2014AMM} propose a model based on Rational Bézier-Bernstein Polynomials for simulating ECG waves. This model utilises Rational Bézier-Bernstein Curves of degree n (see eq \ref{eq1} \cite{Chutchavong2014AMM}), which can represent a variety of shapes, including circles, ellipses, parabolas, and hyperbolas \cite{Chutchavong2014AMM}. By maintaining the Rational Bézier-Bernstein Curves as the foundation and modifying other aspects of the model, our proposed model is used to impute HR values effectively. The model's adaptability stems from the extraction of HR values from ECG data using preprocessing techniques.

\begin{equation} \label{eq1}
R(t) = \frac{\sum_{i=0}^n p_i w_i B_i^n(t)}{\sum_{i=0}^n w_i B_i^n(t)}
\end{equation}
Here $B_i ^{n}(t)$ is the Bernstein polynomial of degree n represented by:
$B_i^n(t) = \binom{n}{i} t^i (1-t)^{n-i}$
and  $w_i$ are the weights corresponding to the control points
($p_i$) for i = 0,1,…,n 
To efficiently apply this model for imputing HR values, the following design is employed: the endpoints of the missing data segments are leveraged, whereby five values from the beginning and five values from the end are filled linearly. This preserves the linear start and end of the data so the imputation procedure starts and ends at the starting and final known value. The intermediate section is then imputed using the Rational Bézier-Bernstein polynomial. To impute the middle section, the surrounding points near the missing gaps, referred as control points, are utilised \cite{s23031454}. Control points are used to capture the HR pattern in the known preceding phase and succeeding phase of the missing values, particularly the peaks and valleys. For example, in a 60-second gap, 30 preceding and 30 following known values serve as control points to fill the missing section. In these control points, extremums (peaks and valleys) receive additional weights (1.5) than other points (1). The maximum number of control points considered is 900, applicable for time intervals of both 900 seconds and 1800 seconds. Including more than 900 control points can lead to overfitting, making the curve excessively sensitive to small changes in control points or weights, resulting in unstable imputed values. Lastly, the final imputed values are constrained within the range of 40 to 160 bpm, as this range is clinically relevant. 

\subsection{\textbf{Controlled PCHIP with Mapped Peak and Valleys of Control Points (CMPV)}}

This method presents an innovative approach to effictively capture the intricate patterns in control points found in HR data. By strategically aligning the peaks and valleys of these control points, the method establishes a robust baseline for imputing missing data gaps. By taking this alignment into account, the imputation of missing values reflects the fluctuations, peaks, and valleys learned from the phases of the control points.
The preliminary phase of linearly imputing five values from the start and five from the end is analogous. This method employs control points similar to the aforementioned technique; however, rather than assigning weights to the extremums, it inverts and maps the preceding peaks and valleys of the control point to correspond to the first half of the imputation region, while similarly inverting the succeeding peaks and valleys of control points to map them to the second half. This inversion is crucial for maintaining continuity, as it ensures that the last part of the preceding control points informs the first part of the missing segment. In real-life scenarios, if a person's heart rate is in a mild phase just before the data gap, our imputation technique learns from the control points' peaks and valleys to precisely fill the starting values of the imputation region with mild phase data. As a result, this mapping creates a template that effectively captures the characteristics of surrounding HR values, facilitating a more accurate fill for the gaps. To complete the pattern of imputation technique, connection between the mapped control points is done by PCHIP polynomials, as referenced in equation \(\ref{eq2}\). At last the clipping of imputed values range is done from 40 to 160 bpm, as this range is clinically pertinent. 
\begin{equation} \label{eq2}
y(x)=\sum_{i=1}^{n-1}(a_i(x-x_i)^3+b_i(x-x_i)^2+c_i(x-x_i)+d_i)
\end{equation}

\subsection{\textbf{Comparison Baselines}}

To compare the proposed imputation approaches, the results are evaluated against the following imputation techniques. These are selected due to their prevalent application in imputing missing HR and ECG measurements in hypoglycemia prediction research. 

\begin{itemize}
   
 \item \textbf{Linear:} It fits a straight line utilising the given endpoints of the gap  to compute the missing values.
 
 \item  \textbf{PCHIP}: This interpolation technique is particularly useful for preserving trends in the data and avoiding overshooting. (see to equation \ref{eq2})

 \item \textbf{K Nearest Neighbor (KNN):}  This imputation method leverages the concept of proximity to estimate missing values. It calculates the distance to all known values for each missing value, selects the closest five neighbours, and imputes the missing value as the mean of these neighbours. This process is repeated for all missing values, ensuring localized and context-aware imputation.
    
 \item \textbf{B Spline: \cite{Mochurad2023ParallelAF}} The process of B-Spline interpolation involves constructing B-Spline basis functions 
\( B_{i,k}(x) \), where \( k \) is the degree, satisfying the conditions: 1) Each basis function \( B_{i,k}(x) \) is defined over the interval \([t_i, t_{i+k+1}]\); 2) \( B_{i,k}(x) \) is a polynomial of degree \( k \) within each interval \([t_i, t_{i+1}]\); 3) Each point \( x \) lies within at most \( k+1 \) neighboring basis functions.

\end{itemize}
\subsection{\textbf{Evaluation Metrics}}

An important task in imputing missing data is evaluating the performance of the imputation models. The goal of imputation is not merely to predict the missing data but to capture the underlying properties of the dataset, such as variability and distribution~\cite{Boursalie2022}. This is especially relevant for  physiological data like HR since unlike a machine, the human body has a lot more variations due to  external factors like stress and physical exercise. Therefore, using a single predictive accuracy metric to compare the effectiveness of different imputation techniques can be misleading. For instance, the RMSE directly compares the imputed values with the actual values, rather than assessing the distributions~\cite{Boursalie2022}. Different metrics capture various aspects of perceptual alignment, which can sometimes lead to contradictory results~\cite{ahlert}. For example, a model that performs well according to one metric may score poorly on another. Therefore, it is essential to explore aggregations of different metrics~\cite{ahlert}. 

Therefore, RMSE~\cite{Boursalie2022}, alongside the following new evaluation metrics, are used and proposed to assess the effectiveness of the imputation methods.
Finally, we have aggregated the scores of the three metrics for overall comparison of the imputation methods.

\textbf{Extremum Density Metric (EDM):}
We have introduced and implemented a novel metric to evaluate the data's fluctuations, especially regarding HR features marked by frequent peaks and valleys. Comprehending this fluctuation is essential, as traditional imputation methods, like spline or polynomial interpolation, sometimes fail to adequately capture the intrinsic variations in HR data, which may lead to misleading RMSE metrics. To evaluate the effectiveness of an imputation technique, we propose the Extremum Density Metric (EDM) (\ref{eq 5}), which measures pattern similarity between the original and imputed datasets by analyzing the extremum density (ED) -- essentially, the frequency of peaks and valleys present in each dataset. The extremum density function computes the number of peaks and valleys across the entire data segment and normalizes this count by the segment length. This evaluation is conducted for both the original and imputed data, subsequently calculating the absolute score difference between the two values (\ref{eq 4}) .
\begin{equation}
\label{eq 4}
\textbf{EDM} = \frac{{\text{NP} + \text{NV}}} {\text{LS}}   
\end{equation}
where NP is the number of peaks, NV is the number of valleys and LS is length of the segment. 
\begin{equation}
\label{eq 5}
\textbf{EDM Score }= |\text{ED(original)} - \text{ED (imputed)}|
\end{equation}
\textbf{Peak Alignment Score Score (PAS):}
We have proposed this novel metric which evaluates the alignment of peaks between the original HR data and its imputed counterparts. As previously discussed, HR values inherently consist of peaks, and accurately replicating these peaks is crucial for reflecting the original HR data characteristics. A significant misalignment in imputed peaks can lead to erroneous interpretations of an individual's activity during specific time frames, subsequently affecting the predictive accuracy of ML models. For instance, real HR data may exhibit peaks up to 120 bpm, whereas imputed data may peak at 90 bpm.
To assess imputation robustness, peaks are identified in both datasets using the \texttt{find\_peaks} function. The comparison is based on the minimum number of peaks observed in either sequence; if the original has 10 peaks and the imputed 8, only 8 are used for comparison. Absolute differences between corresponding peak values are calculated, and the PAS score is derived as their mean (\ref{eq 6}). If no peaks are detected in either sequence, a default score of 15 is assigned to prevent computational errors, determined as 0.48 above the highest observed PAS value (14.52). This method not only quantifies the alignment but also provides critical insight into the fidelity of the imputation process, thereby influencing the overall performance of subsequent analyses and predictive modeling efforts.

\begin{equation}
\label{eq 6}
\textbf{PAS} = \text{average} \left( \left| y_{\text{peak (original)}} - y_{\text{peak (imputed)}} \right| \right)
\end{equation}
The metric scores for each time interval are averaged based on the number of gaps identified. For instance, in the 30-second interval, 65 gaps are detected. The overall metric scores for that time span are calculated by averaging the metric scores for each of the 65 gaps.

\textbf{Combined Metric (CM):}
Aggregation of the scores of different metrics can be perceived as a multidimensional concept, in which different scores measure different aspects of the datasets~\cite{ahlert, gupta, framshap}. For combining and giving equal weights to the individual metric we have used the aggregation method derived from the paper of Gupta et al~\cite{gupta, framshap}. 
To calculate the combined metric score for each time period, we first normalize the scores from each metric to make them easier to compare and to bring them in a unified scale. We determine the normalized scores for the three metrics for each time gap in the data based on metrics scores of different imputation techniques. The combined score formulates with these assigned weights:
$\text{weight}_{\text{RMSE}} (\alpha)$  = 0.333, $\text{weight}_{\text{EDM}} (\beta)$ = 0.333, $\text{weight}_{\text{PAS}} (\gamma)$ = 0.333
\begin{equation}
\label{eq5}
\textbf{CM} = \alpha * RMSE + \beta * EDM +\gamma * PAS
\end{equation}
Lower individual metric scores indicate better imputation techniques; thus, a lower combined score also signifies a better imputation technique.

\section{Results}
\label{sec:Results}

This section presents the results of the experiments as described in section \ref{table2}, which were conducted using the design outlined above. These experiments hold significant value in the context of imputation techniques.

The lowest number of gaps occurs at 30 seconds and 1800 seconds. For the 30-second time interval, the control points exhibited fewer fluctuations, resulting in fewer peaks and valleys. Meanwhile, for the 1800-second interval, there was insufficient data to utilize additional control points. Overall, an average of approximately 80 gaps were identified for each time interval across the patients in the dataset.

The evaluation metrics analysed imputation techniques for each time interval: 30, 60, 300, 600, 900 and 1800 seconds. When it comes to RMSE, the Linear Imputation technique performs better among all the imputation techniques for all the time intervals. However, as the time interval increases, the RMSE score for all the imputation techniques, including Linear Imputation, also increases. As per the EDM score, CMPV, that we introduce in this paper, consistently outperforms the other imputation techniques. Notably, the performance shows a promising trend of improvement as the length of the time interval increases, showcasing its adaptability. Based on PAS score, CMPV performs the best for 30 seconds and CRBC performs best for 60 seconds; KNN performs the best for all the other time gaps, closely followed by CMPV with approximate average difference of 0.6 from PAS score of KNN. The CM scores indicate that CMPV consistently outperforms all other imputation techniques across various time gaps, with the exception of the 1800 second time interval, where it is surpassed by KNN by approximately difference of 0.04 in the CM scores. Up to the 600 second mark, our other proposed method, CRBC demonstrates the second-best performance; however, for the 900 second gap, KNN ranks as the second-best method.

These results emphasize the reliability of CMPV, establishing it as the most robust imputation technique for short-time gaps of varying lengths. CMPV maintains a stable performance for RMSE metric and PAS metrics for all time intervals and turns out to be better on the basis of EDM scores. CMPV maintains a steady CM score between 0.31 and 0.42 across all time intervals and outperforms other techniques on the basis of CM metric in 5 out of 6 instances. While CRBC exhibits strong performance for time intervals up to 600 seconds for three individual metrics and CM metric, its effectiveness declines for larger gaps starting at 900 seconds. Conversely, KNN underperforms for individual metrics and maintains CM scores values predominantly above 0.5 until the 900 second interval, ultimately outperforming CMPV for the 1800 second gap with a CM score of 0.38. 

\begin{table}[h]
\centering
\caption{Imputation results}
\label{table2}
\setlength{\tabcolsep}{4pt}
\begin{tabular}{|p{30pt}|p{30pt}|p{30pt}|p{30pt}|p{30pt}|p{30pt}|p{30pt}|p{30pt}|}
\hline 
\textbf{Sec} & \textbf{Metric} & \textbf{Linear}  & \textbf{BSpline} & \textbf{KNN} & \textbf{PCHIP} & \textbf{CRBC} & \textbf{CMPV} \\
\hline
30  & CM &0.67  & 0.84 &0.51  & 0.72 & 0.45 & \textbf{0.31} \\
\hline
  N.O.G       & RMSE            &\textbf{4.29} & 5.32 &5.19  & 4.51 & 5.57 & 5.47  \\
       \hline
   65     & EDM             & 0.14 & 0.13 &0.10& 0.14 &0.09 &\textbf{0.07}  \\
       \hline
       & PAS             & 15 &12.91& 8.48 & 15 & 5.54 & \textbf{4.98} \\
       \hline

\hline
60  & CM & 0.66 &0.91 & 0.59 &0.72  &0.35 & \textbf{0.33} \\
\hline
   N.O.G    & RMSE            & \textbf{5.17} & 6.67 & 6.42 &5.45  &6.55  & 6.57 \\
       \hline
   93    & EDM             & 0.14 &0.14  & 0.12 & 0.14 & 0.07 & \textbf{0.06} \\
       \hline
       & PAS               & 15   &12.80 &  7.55&  15.0 & \textbf{5.89} &  6.53\\
\hline

\hline
300  & CM & 0.65 & 0.77 &0.52 & 0.69 & 0.46 & \textbf{0.26 }\\
\hline
    N.O.G    & RMSE           & \textbf{8.27 } &10.45   & 10.39 & 8.64 &11.48  &  10.54\\
     \hline
       101    & EDM             & 0.12 &0.13  &0.12&0.12  &0.05& \textbf{0.05} \\
        \hline
        & PAS               & 15 &12.49& \textbf{8.33}& 15.0 & 10.18 & 8.79 \\
\hline

\hline
600  & CM &0.62  & 0.87 & 0.56 & 0.64 &0.54  & \textbf{0.31} \\
\hline
   N.O.G       & RMSE            &\textbf{10.50 }&13.22  & 13.14 & 10.69 &13.61& 13.22 \\
        \hline
  87       & EDM             &0.12  & 0.13 & 0.12 & 0.12 &0.08  & \textbf{0.04 }\\
        \hline
        & PAS               & 15.0 &13.62  & \textbf{9.51} &15.0  & 10.54 & 9.80 \\
        \hline

\hline
900  & CM & 0.64 & 0.77 &0.51  & 0.69 &  0.61& \textbf{0.34}  \\
\hline
 N.O.G        & RMSE            & \textbf{10.81} &12.90  & 12.84 & 11.35 & 14.09 & 13.71 \\
        \hline
  81      & EDM             & 0.12 & 0.13 &0.12&0.12  & 0.09 & \textbf{0.04} \\
        \hline
        & PAS              &  15&13.25  & \textbf{9.82}  & 15 &11.49  & 10.53 \\
        \hline

\hline
1800  & CM &0.64  &0.60  & \textbf{0.38} &0.68  & 0.46 & 0.42 \\
\hline
 N.O.G       & RMSE           &  \textbf{10.35} & 10.91 &10.83&10.60  &11.26  & 12.39 \\
         \hline
   48      & EDM             & 0.12 &0.13  & 0.12 & 0.12 &0.11  & \textbf{0.04} \\
         \hline
         & PAS              &15& 11.70 & \textbf{8.12 } &8.58  &9.63  & 9.86 \\
\hline

\end{tabular}
\end{table}

\section{Discussion}
\label{sec:Discussion}

\subsection{Rationale of Using Imputation Techniques}
The effectiveness of imputation techniques is crucial for preserving natural and informative patterns in time-series data. Missing values in datasets are pertinent and are often the result of human error or sensor failures. Shorter gaps, ranging from 30 seconds to 30 minutes, are more common than larger gaps and can arise due to various factors, such as synchronization errors, battery limitations, enviromental interference, connectivity issues or the removal of outliers for brief intervals. These small gaps can lead to the loss of critical data, including extreme events such as abnormal heart rate or glucose fluctuations. This study addressed these limitations by imputing missing data, thereby enhancing the overall data quality and improving the predictive performance of extreme physiological states and enhancing the accuracy of machine learning models for hypoglycaemia prediction.

\subsection{Temporal Variability of HR values }
HR values naturally fluctuate due to various external factors, even in steady states, in which some variability persists. The HR trend is inherently influenced by both preceding and subsequent phases—referred to as control points in our experimental setup— whose values can be utilised for imputing missing gaps. For example, if a person is doing physical exercise or is experiencing a steady fall in BG values, the subsequent phases around the missing values can be utilised to predict and impute the missing gap. However, existing popular short-interval imputation methods often disregard surrounding phases, reducing their accuracy. Linear interpolation connects only the first and last known points with a straight line, while PCHIP and B-spline methods use piecewise polynomials but fail to preserve the HR pattern. KNN relies on neighbouring values but does not fully capture phase continuity. This study compares state-of-the-art imputation techniques with our proposed methods. Machine or deep learning-based approaches were excluded from the baseline comparison, as they are more suitable for long-term imputation (gaps over 2 hours) and require extensive training data~\cite{SuwenLin, Mochurad2023ParallelAF}.

\subsection{Proposed Imputation Techniques and Evaluation Metrics}

Our proposed methods, CRBC and CMPV, are designed with an understanding of the dynamics and functioning of the human heart. These approaches leverage available HR data to identify and reflect the phases preceding and following a missing segment, enabling the accurate prediction of missing values. HR variations are influenced by the individual's activity levels, with fluctuations characterized by peaks during periods of exertion and valleys during rest phases. Recognizing this inherent pattern, our methods prioritize these fluctuations to enhance the imputation process. CRBC utilizes control points by assigning weights to the peaks and valleys, ensuring a more precise reconstruction of missing values. In contrast, CMPV employs an inverse mapping approach, systematically transferring the peak and valley structures of control points onto the missing regions. Moreover, our methods can facilitate diagnosis by identifying patterns in missing values based on surrounding data. In cardiovascular monitoring, if imputed data aligns with physiological trends while recorded values deviate—exhibiting irregular rhythms, erratic fluctuations, or abnormal heart rates—this discrepancy may indicate conditions such as arrhythmias or cardiac stress. 

To assess the effectiveness of imputation techniques, the existing literature primarily employs metrics such as RMSE, MAPE, and MAE, which focus on comparing imputed estimates to actual values rather than evaluating their distributions~\cite{Boursalie2022}. To fill this critical gap, we present two metrics: EDM and PAS. EDM analyzes the frequency density of imputed values relative to the original data, while PAS assesses the alignment of peaks between imputed and actual values. 
These metrics are designed considering the physiological characteristics of the heart. To evaluate how accurately the imputed values preserve the intricate patterns of heart rate fluctuations, an aggregated metric is employed for comprehensive assessment. This unified measure considers multiple facets of the original data, including proximity to missing values and distribution characteristics, to determine the best imputation method.

\subsection{Validation of Methods}

The performance of the proposed imputation techniques is analyzed based on individual evaluation metrics as well as an aggregated metric, the CM score. Linear imputation performs the best on the basis of RMSE scores, our proposed method CMPV performs the best for EDM score and KNN demonstrates a superior performance in 4 out of 6 time periods. The proposed methods in this study, CRBC and CMPV, have a comparable performace on the basis of RMSE score with respect to linear imputation with an average score of 10.42 and 10.32 respectively, while linear has 8.23. On the basis of the PAS score, CMPV demonstrates stability and reliability in all time gaps with an approximate average difference of 0.6 from KNN.
This understanding of different evaluation metrics favouring different imputation techniques validates our claim that single evaluation metric can lead to misleading results. On the basis of the CM score, CMPV outperforms other methods 5 out of 6 times, showcasing its stable performance throughout all the individual metrics and CM score.

The average CM values across all time gaps reinforce the enhanced performance of CMPV and CRBC, demonstrating a significant performance difference. Our methods yield average CM scores of 0.48 for CRBC and 0.33 for CMPV, marking a substantial improvement with the next best score of 0.52 for KNN and over other traditional methods, respectively. In summary, the advantages of our proposed methods lie in their ability to capture the intricate dynamics of the heart, resulting in a more accurate imputation process. 
 
\subsection{Personalised Heart Rate Simulator}
This study represents a foundational step toward developing a personalised HR simulator capable of forecasting heart rate dynamics based on the patient's physiological phases. Such a model could function as a standalone predictive tool and contribute to early intervention strategies for medical conditions beyond hypoglycemia that are linked to heart rate variability. Additionally, this approach has the potential to improve effectiveness of wearable health devices by enabling real-time anomaly detection and issuing early warnings for extreme HR fluctuations, ultimately improving patient monitoring and preventive healthcare measures. 

\section{Conclusion}
\label{sec:Conclusion}

This study highlights the importance of imputing missing data to enhance the performance of detection of extreme events in HR and BG values. The practical implications of our research are significant, as it provides a more accurate and reliable method for handling missing HR values. Our proposed imputation techniques, CRBC and CMPV, outperform the traditional imputation techniques for imputing HR values over a short-term horizon. By assessing the effectiveness of the imputation methods using various metrics -- including one state-of-the-art metric, two proposed novel metrics, and one proposed combined metric -- we demonstrate that our methods lead to a significant improvement in accuracy. Specifically, we achieve an average improvement of approximately 4\% for the CRBF method and 19\% for the CMPV method compared to state-of-the-art imputation techniques, as measured by the combined metric across all time gaps.

Our work in addressing the limitations of missing values in HR features is a fundamental step towards improving the data quality and supporting more accurate prediction models. This research not only contributes to the advancement of imputation techniques but also provides a strong foundation for the development of a heart rate simulator for forecasting HR values with the aid of real-time data. 

\subsection{Limitation and Future Work}
Currently, 30 seconds to 30 minutes of HR data have been imputed, with performance declining as time horizons increase. Future work will focus on improving accuracy by optimising weights and control points, a limitation of this study due to the small dataset size. Notably, parameter fine-tuning  requires different datasets representing a variable population. Moreover, an enhanced framework aggregating different evaluation metrics into a unified structure supports a more robust evaluation that can be used in different domains.

To ensure the generalizability and data independence of our framework, we will apply it to diverse datasets, evaluating its effectiveness in improving short-term detection accuracy for critical conditions such as hypoglycemia in patients with T1D patients. Furthermore, we will explore optimal imputation strategies for extended time horizons, advancing a multi-step, multi-model imputation framework specifically designed for HR data. Also, a hybrid imputation technique will be explored utilising the framework of Physics Informed neural Networks (PINNs) architecture~\cite{Gupta2025PINN_SHM}. These efforts will pave the way for more reliable, data-driven clinical insights, ultimately enhancing early detection and intervention for critical health conditions.

\bibliographystyle{ieeetr}
\bibliography{literature.bib}

\end{document}